\documentclass[final,5p,times,twocolumn]{elsarticle}

\usepackage{amsmath}
\usepackage{amssymb}
\usepackage{amsfonts}
\usepackage{graphicx}

\biboptions{square,sort&compress}

\newcommand{\pp}{\boldsymbol{p}}
\newcommand{\qq}{\boldsymbol{q}}

\newcommand{\free}{\text{(free)}}
\newcommand{\scat}{\text{(scat)}}

\newcommand{\dd}{\text{d}}

\newcommand{\nppp}{\nonumber\\[3pt]}
\newcommand{\pppp}{\\[6pt]}

\newcommand{\rr}{\boldsymbol{r}}

\newcommand{\kk}{\boldsymbol{k}}
\newcommand{\LL}{{\mathcal L}}

\newcommand{\hc}{\text{H.c.}}
\newcommand{\bra}[1]{\langle#1|}

\newcommand{\ket}[1]{|#1\rangle}

\newcommand{\Fig}[1]{Fig.~\ref{#1}}

\newcommand{\Ref}[1]{Ref.~\cite{#1}}
\newcommand{\Refs}[1]{Refs.~\cite{#1}}
\newcommand{\refs}[1]{~\cite{#1}}
\newcommand{\eq}[1]{~\eqref{#1}}

\newcommand{\Eq}[1]{Eq.~\eqref{#1}}
\newcommand{\Eqq}[2]{Eqs.~\eqref{#1} and \eqref{#2}}

\begin{document}

\begin{frontmatter}

\title{Cosmic quantum optical probing of quantum gravity through a gravitational lens}

\author{Charles H.-T. Wang}
\ead{c.wang@abdn.ac.uk}

\author{Elliott Mansfield}

\author{Teodora Oniga}

\address{Department of Physics, University of Aberdeen, King's College, Aberdeen AB24 3UE, United Kingdom}

\begin{abstract}
We consider the nonunitary quantum dynamics of neutral massless scalar particles used to model photons around a massive gravitational lens. The gravitational interaction between the lensing mass and asymptotically free particles is described by their second-quantized scattering wavefunctions. Remarkably, the zero-point spacetime fluctuations can induce significant decoherence of the scattered states with spontaneous emission of gravitons, thereby reducing the particles' coherence as well as energy. This new effect suggests that, when photon polarizations are negligible, such quantum gravity phenomena could lead to measurable anomalous redshift of recently studied astrophysical lasers through a gravitational lens in the range of black holes and galaxy clusters.
\end{abstract}


\end{frontmatter}

% \linenumbers

\section{Introduction}

%\textbf{Introduction.}
The effort to find a quantum theory of gravity remains as a key objective of theoretical physics in the 21st century. To better guide theoretical development, there has been substantial interest in searching for possible experimental indications\refs{Amelino2013, Berti2015}. %Unfortunately, this too has had limited success.

There exist a number of important studies that could lead to a future detection of quantum gravity effects. Recently, the data from the first detection of gravitational waves by LIGO\refs{LIGO2016} have been analyzed further to search for echoes that would indicate the Planck-scale structure that some theories predict would exist on the event horizon of a black hole\refs{Cardoso2016}. Gravitational wave astronomy is a promising prospect in the effort to sense quantum gravity and there have been recent developments suggesting coherent Rydberg atoms could be sensitive to the zero-point spacetime fluctuations and stochastic gravitational waves\refs{Quinones2017}.

Attempts to probe quantum gravity can go back as far as 1981, when an argument
against semiclassical relativity was given in \Ref{Page1981}, claiming indirect evidence of quantum gravity. While this may be an incomplete dichotomy, as there are other theories that attempt to ``fix'' gravity, such as Modified Newtonian Dynamics\refs{Milgrom1983} and Emergent Gravity\refs{Verlinde2016}, it at least reflects the need to quantize gravity as being increasingly recognized.

A few years later, \Ref{Goldman1986} outlined some geophysical experiments that seem to suggest a repulsive term in the full gravitational equation, but this had a large uncertainty. Recently, there have been considerable developments in testing quantum phenomena in the astronomical domain, noticeably including tests for vacuum birefringence\refs{Mignani2017}, quantum nonlocality\refs{Handsteiner2017}, and indeed attempts to observe spacetime foam as a quantum gravity effect\refs{Vasileiou2015}.

Decoherence is known to be a large problematic factor for quantum applications that causes qubits to lose quantum information\refs{Chuang1995, Schlosshauer2014}. Yet, in exploring quantum nature, decoherence can provide useful tools\refs{Pikovski2012, Pfister2016}. Decoherence is an essential part of open quantum dynamics, where dissipative interactions with the environment lead to decays from excited states towards the ground state in systems such as atoms in quantum optics\refs{Schlosshauer2014, Breuer2002}. Recently, much work has been carried out in regard to gravitational decoherence with more general systems\refs{Quinones2017, Oniga2016a, Oniga2016b, Oniga2016c, Oniga2017a}.

Any cosmic search for quantum gravitational decoherence are constrained by sources not emitting coherent radiation. While pulsars could be potentially considered, since they do not emit coherently in the optical spectrum, their tests for gravitational decoherence using modern quantum optics techniques may be inviable.

Fortunately, celestial objects known as astrophysical lasers have been studied in more recent times\refs{Dravins2007, Johansson2004}. These sources naturally induce a population inversion by one of two methods: (i) the rapid cooling of ionized plasma, causing electronic recombination, and (ii) selective radiative excitation, where emission lines of similar wavelengths may induce the transition to a higher state\refs{Dravins2007, Johansson2004}.

Therefore astrophysical lasers appear to be suitable radiation sources for cosmic probing of gravitational decoherence. However, it has been pointed out that photons and other particles in free space are insensitive to spacetime fluctuations\refs{Oniga2016b}. While radiation can undergo gravitational decoherence in a cavity, the size of the effects are negligible in normal laboratory scales that limit the quadrupole moments of the quantum system, which provide the effective coupling to gravitational fluctuations\refs{Oniga2016c}.

In view of these recent developments, here we show that gravitationally lensed coherent optical radiations on the astronomical scale could emerge as a viable test bed for quantum gravity through decoherence and the resulting anomalous redshift. We consider astrophysical lasers to provide the coherent optical radiation sources and objects ranging from black holes to galaxy clusters as lensing  masses.

We work in the interaction picture with the relativistic units where $c=1$ unless otherwise stated. The framework of quantized linearized gravity is adopted with metric signature $(-1,1,1,1)$ and transverse-traceless (TT) coordinates.

%$\lambda_*=$ wavelength of light

%$f_*=$ frequency of light

%~\\

%\newpage
\section{Open quantum dynamics of a gravitational lensing system}
%We consider $D \gg L=1/\epsilon \gg d_* \gg \lambda_*$.
To focus on the gravitational decoherence and the resulting dissipative phenomena of gravitationally lensed light, here we will neglect self gravity, environmental gravitational temperature, and the spin polarizations of the photons and model them with a neutral massless scalar field $\phi(\rr,t)$ subject to an external potential represented by $u(\rr)$ described by the Lagrangian density
\begin{eqnarray}
\LL
&=&
-
\frac12\,
\eta^{\mu\nu}\phi_{,\mu}\phi_{,\nu}
-u\, \phi^2.
\label{lag}
\end{eqnarray}
Writing
$\phi=\psi(\rr)\,e^{-i\omega t} + \hc$,
we see that the field equation\eq{lag} is satisfied by an effective time independent Schr\"odinger equation for $\psi(\rr)$.

We consider a gravitational potential
$
\Phi=-G M {e^{-\epsilon r}}/{r}
$
where $M$ is the lensing mass and $\epsilon$ is a small Yukawa parameter to regularize the infinite range of the Newtonian potential.
%The gravitational refractive index $n=1-2\Phi$ then gives raise to the dispersion relation
%\begin{eqnarray}
%\omega
%&=&
%(1+2\Phi)p
%\end{eqnarray}
%
%Here we denote wave vectors associated with the scalar field by $\pp, \qq, \cdots$ and wave vectors associated with the gravitons by $\kk, \cdots$
%
Here we consider the wave vector $\pp$ of the scalar field to have a characteristic norm $p = |\pp| \sim p_*$. This allows us to approximate the potential function by
\begin{eqnarray*}
u(\rr)
&=&
-2\,p_*^2 G M \frac{e^{-\epsilon r}}{r}.
\end{eqnarray*}
%This corresponds
%\begin{eqnarray}
%V(r)
%&=&
%-\frac{2\hbar^2 G M p_*^2}{m\,r}
%\end{eqnarray}
%
The effective time independent Schr\"odinger equation for $\psi(\rr)$ can then be solved in terms of the Fourier transform
\begin{eqnarray*}
%\psi_{\qq}(\pp)
%&=&
%\frac1{(2\pi)^3}\int\dd^3r\,e^{-i\pp\cdot\rr}\psi_{\qq}(\rr)
%\label{r2p}
%\ppp
\psi_{\qq}(\rr)
&=&
\int\psi_{\qq}(\pp)\,e^{i\pp\cdot\rr}\dd^3p
%\label{p2r}
\end{eqnarray*}
by the approximate expansion
\begin{eqnarray*}
\psi_{\qq}(\pp)
&=&
\delta(\pp-\qq) +
%\nppp&&
\frac{2\, G M p_*^2}{\pi^2[p^2-(q+i\epsilon)^2][|\pp-\qq|^2+\epsilon^2]}
%\label{psikkppy}
\end{eqnarray*}
analogous to the momentum representation of the Coulomb scattering wavefunctions\refs{Guth1951, Wong2004}.
The first term above represents the plane wave associated with the asymptotically free particle and the second term represents the scattered spherical wave due to gravitational lensing.
The full scalar field operator then takes the form
\begin{eqnarray}
\phi
&=&
\int\!\dd^3q
\sqrt{\frac{\hbar}{2(2\pi)^3{q}}}\;
a_{\qq}\psi_{\qq}(\rr)\,e^{-i q\, t}
+
\hc
\label{phiex}
\end{eqnarray}
in terms of the annihilation operator $a_{\qq}$.

Under the Markov approximation which neglects short-term memory effects\refs{Breuer2002}, the general gravitational master equation recently obtained in \Ref{Oniga2016a} yields
\begin{eqnarray}
\dot\rho(t)
&=&
-
\frac{8\pi G}{\hbar}
\int\! \frac{\dd^3 k}{2(2\pi)^3k}
\int_{0}^{\infty} \dd s\,
e^{-i k s}
\nppp&&
\big\{
[
\tau^\dag_{ij} (\kk,t),\,
\tau_{ij}(\kk,t-s) \rho(t)
]
+ \hc
\big\}.
\label{maseqnmkc}
\end{eqnarray}
In the adopted TT coordinates, spacetime fluctuations generally induce a fluctuating correction to the potential and hence $\tau_{ij}$, which is significant for a harmonic potential\refs{Oniga2017a}. However, for the present gravitational potential, such corrections are negligible.

The Fourier-transformed TT stress tensor $\tau_{ij}(\kk,t)$ that provides the coupling between the particles and spacetime fluctuations using \Eq{maseqnmkc} follows from \Eqq{lag}{phiex} to be
\begin{eqnarray}
\tau_{ij}(\kk,t)
&=&
\int\dd^3p'\dd^3p''\,
\Big[
\tau_{ij}(\kk,\pp',\pp'')\,
e^{-i({p'}-{p''}) t}
\nppp&&
+
\tau_{ij}^\dag(-\kk,\pp',\pp'')\,
e^{i({p'}-{p''}) t}
\Big]
\label{tkksc}
\end{eqnarray}
where
\begin{eqnarray*}
\tau_{ij}(\kk,\pp,\pp')
&=&
\tau_{ij}^\free(\kk,\pp,\pp')
+
\tau_{ij}^\scat(\kk,\pp,\pp')
%\label{tkpp}
\end{eqnarray*}
in terms of
\begin{eqnarray*}
%\hspace{-10pt}
\tau_{ij}^\free(\kk,\pp,\pp')
&=&
\frac{\hbar}{2}\,
\frac{a_{\pp'}^\dag a_{\pp}\,P_{ijkl}(\kk) p_k p_l}{\sqrt{{p}{p'}}}\,
\delta(\pp-\pp'-\kk)
%\label{tkpp0}
\pppp
%\hspace{-10pt}
\tau_{ij}^\scat(\kk,\pp,\pp')
&=&
\frac{\hbar\, G M p_*^2}{\pi^2}\,
\frac{a_{\pp'}^\dag a_{\pp}\,P_{ijkl}(\kk)}
{\sqrt{{p}{p'}}[|\pp-\pp'-\kk|^2+\epsilon^2]}\,\times
\nppp&&
\hspace{-45pt}
\Big\{
\frac{p_k p_l}{|\pp-\kk|^2-(p'-i\epsilon)^2}
+
\frac{p'_k p'_l}{|\pp'+\kk|^2-(p+i\epsilon)^2}
\Big\}.
%\label{tkpp1}
\end{eqnarray*}
Here the first term $\tau_{ij}^\free(\kk,\pp,\pp')$ is associated with the asymptotically free particle states which does not contribute to the master equation\eq{maseqnmkc} as discussed in \Ref{Oniga2016b}, whereas the second term $\tau_{ij}^\scat(\kk,\pp,\pp')$ contains the scattered spherical waves and contributes to \Eq{maseqnmkc} with decoherence effects.

Considering the gravitational interaction to be weak, we assume the density matrix of the particles to evolve perturbatively over time so that
\begin{eqnarray*}
\rho(t)=\rho_0+\Delta\rho(t)
\end{eqnarray*}
with $|\rho_0| > |\Delta\rho(t)|$ where $\rho_0=\rho(t=0)$. Then by substituting \Eq{tkksc} into \Eq{maseqnmkc}, taking the limit $t\to\infty$, i.e. $t \gg 1/p_*$, applying the Sokhotski-Plemelj theorem\refs{Oniga2016a} and neglecting the nondissipative Lamb-Stark shift terms, and repetitively using the same argument leading to free particles suffering no Markovian gravitational decoherence\refs{Oniga2016b}, we obtain the asymptotic change of the density matrix of the form
\begin{eqnarray}
\Delta\rho
&=&
%-\frac{\hbar\kappa\nu^2}{32\pi^4}
-\frac{\zeta}{2}
\int \dd^3 k\,\dd^3 p\,\dd^3 q\,\dd^3 p'\,\dd^3 q'
%\nppp&&
\nppp&&
\hspace{-5pt}\times\,
\frac{
\delta(k-{p}+{p'})
\delta(k+{q}-{q'})
P_{ijkl}(\kk)}
{k\sqrt{{p}{p'}{q}{q'}}\,
[|\pp-\pp'-\kk|^2+\epsilon^2][|\qq-\qq'+\kk|^2+\epsilon^2]}
\nppp&&
\hspace{-5pt}\times\,
\Big\{
\frac{p_i p_j}{|\pp-\kk|^2-(p'-i\epsilon)^2}
+
\frac{p'_i p'_j}{|\pp'+\kk|^2-(p+i\epsilon)^2}
\Big\}
\nppp&&
\hspace{-5pt}\times\,
\Big\{
\frac{q_k q_l}{|\qq+\kk|^2-(q'+i\epsilon)^2}
+
\frac{q'_k q'_l}{|\qq'-\kk|^2-(q-i\epsilon)^2}
\Big\}
\nppp&&
\hspace{-5pt}\times\,
\Big\{
a_{\qq'}^\dag a_{\qq}
a_{\pp'}^\dag a_{\pp}\,\rho_0
-
a_{\pp'}^\dag a_{\pp}\,\rho_0
a_{\qq'}^\dag a_{\qq}
\Big\}
+\hc
\label{maseqnmkf10}
\end{eqnarray}
with $\Delta\rho=\Delta\rho(t\to\infty)$ and a dimensionless parameter
\begin{eqnarray}
\zeta
&=&
\frac{32 h\, G^3 M^2 f_*^4}{c^{10}}.
\label{Gamm}
\end{eqnarray}
In the above, the speed of light $c$ has been restored, $h$ is the Planck constant, and $f_*=c p_*/2\pi$ is the frequency of the particle. As will be clear below, since this frequency is subject to a redshift due to quantum dissipation of the particle when scattered by the lensing mass determined by $\zeta$, we will refer to it as the ``redshift coefficient''.

%\bra{}\rho_0\ket{}
\begin{figure}[!tbp]
%\vspace{5pt}
\includegraphics[width=1\linewidth]{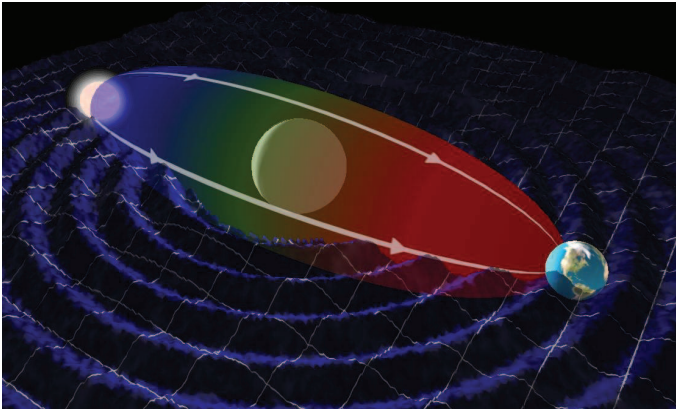}
\vspace{17pt}
\caption{{\bf An illustration for quantum gravitational redshift of lensed coherent light through spontaneous emission of gravitational waves induced by spacetime fluctuations.} The nonunitary dynamics of coherent light modelled as massless scalar field experience redshift through gravitational lensing, represented by the blue-to-red colour varying surface, extending from the bright source in the far view, around a lensing mass represented by the central sphere, to the Earth in the near view.
The background navy curved surface illustrates the outwards propagating gravitational waves spontaneously emitted by the lensed light superimposed with random ripples indicating spacetime fluctuations due to zero-point quantum gravity.}
\label{fig_lens}
\vspace{15pt}
\end{figure}

%~\\

\section{Quantum gravitational redshift through gravitational lensing}

To understand the implication of \Eq{maseqnmkf10}, let us start with a relatively simple one-particle case. Then the corresponding matrix elements read
\begin{eqnarray}
\hspace{-10pt}
\bra{\pp_2}\Delta\rho\ket{\pp_1}
&=&
\frac{\zeta}{2}\int\dd p
\int\dd^3\Omega\,
\Big\{
F_1\,\bra{\qq}\rho_0\ket{\pp_1}
\big|_{\substack{{q}=p_2\\k=-{p}+p_2}}
\nppp
&&
%\hspace{-50pt}
+
F_2\,\bra{\pp}\rho_0\ket{\qq}
\big|_{\substack{{q}={p}+p_1-p_2\\k={p}-p_2}}
\Big\}
+\hc
\label{k2rhok1a}
\end{eqnarray}
%subject to
%\begin{eqnarray}
%k > 0
%\label{kpqcnd}
%\end{eqnarray}
where
$
\int\dd^3\Omega
=
\int\dd\Omega(\pp)\dd\Omega(\qq)\dd\Omega(\kk)
$
and $F_1$ and $F_2$ are certain expressions of vectors  $\pp_1,\pp_2,\pp,\qq,\kk$ derived from \Eq{maseqnmkf10} with similar constructions.
%\begin{eqnarray}
%F_1 = 0 \hbox{ for }  p > p_2,
%\quad
%F_2 = 0 \hbox{ for }  p < p_2
%\label{F12}
%\end{eqnarray}
An isotropic monochromatic radiation source can be modelled by the initial density matrix
\begin{eqnarray}
\bra{\pp}\rho_0\ket{\pp'}
&=&
\frac{1}{4 \pi p_*^2}\,
\sqrt{\delta(p-p_*)\delta(p'-p_*)}.
\label{Sro3}
\end{eqnarray}

By evaluating \Eq{k2rhok1a} with this initial density matrix\eq{Sro3}, we find that the scattered states contain broadened frequencies no higher than the initial frequency as {$\bra{\pp_2}\Delta\rho\ket{\pp_1}=0$} for $p_1 > p_*$ or $p_2 > p_*$.
It also follows that {$\bra{\pp_2}\Delta\rho\ket{\pp_1}=0$}
for $p_1 \neq p_2$ and hence there is no superposition between different redshifted frequencies in the scattered states.

Therefore the nonunitary dynamics of the particles experience redshift through gravitational lensing as a result of quantum dissipation.
Physically, as illustrated in \Fig{fig_lens}, the loss of energy is induced by the spacetime granulation of scattered states having different gravitational energies with the lensing mass causing spontaneous decay of states accompanied by spontaneous emission of gravitons.
This process can be understood as the gravitational analogue of the electromagnetic spontaneous decay of an electron with spontaneous emission of photons in an atom.

\begin{figure}[!tbp]
%\vspace{5pt}
\includegraphics[width=1\linewidth]{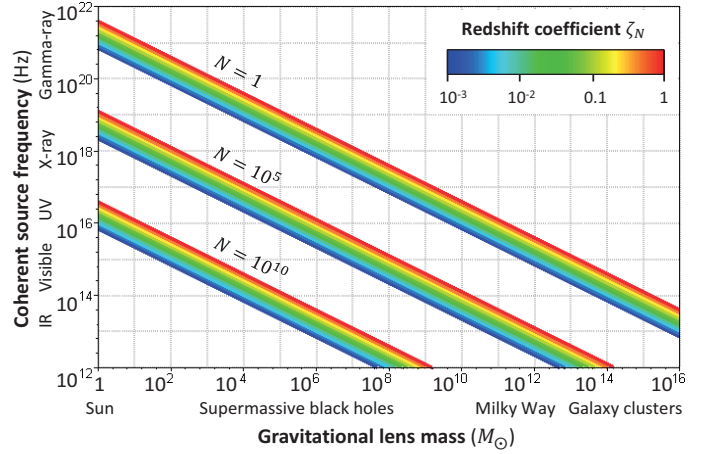}
\vspace{0pt}
\caption{{\bf Quantum gravitational redshift of coherent astronomical radiation at different frequencies in the presence of a gravitational lens for a range of mass scales.} As shown, for optical frequencies at $10^{14}$--$\;10^{15}$~Hz, the redshift coefficient $\zeta_N$ can come close to order one with a galaxy cluster mass for $N=1$ to a supermassive black hole mass for $N=10^{10}$.
As a result, the effect could lead to measurable anomalous redshift of gravitationally lensed radiations from astrophysical lasers as observational evidence of quantum gravity.}
\label{fig_shift}
\vspace{15pt}
\end{figure}

The total probability of transition, i.e. decay ratio, into to redshifted states can be evaluated by tracing $\Delta\rho$ as follows
\begin{eqnarray*}
\int_{p<p_*}\!\bra{\pp}\Delta\rho\ket{\pp}\,\dd^3\pp
\,\Big|_{\epsilon \to 0}
%\nppp
%&
\approx%&
\zeta
%\label{k2ok1f4}
\end{eqnarray*}
up to an order one numerical factor. This justifies $\zeta$ as the one-particle redshift coefficient through the described quantum gravitational lensing mechanism.

For $N$ particles, following the collective quantum analysis in \Refs{Quinones2017, Oniga2016b, Oniga2016c, Oniga2017a}, the gravitational decoherence of $N$ bosonic particles occupying the same state generally scales with $N^2$, by virtue of the quadratic dependence of the gravitational master equation on $\tau_{ij}$ as in \Eq{maseqnmkc}. We therefore introduce the effective redshift coefficient $\zeta_N = \zeta\,N^2$ for coherent $N$-particle states. Using \Eq{Gamm}, we can evaluate $\zeta_N$ for different frequencies, lensing masses and particles numbers.

As shown in \Fig{fig_shift}, for optical frequencies at $10^{14}$--$\;10^{15}$~Hz, the redshift coefficient $\zeta_N$ can come close to order one with a galaxy cluster mass for $N=1$ to a supermassive black hole mass for $N=10^{10}$.
This could lead to measurable anomalous redshift of gravitationally lensed radiations from astrophysical lasers as observational evidence of quantum gravity, which may further guide theoretical work.

\section*{Acknowledgments}

The authors are grateful for financial support to the EPSRC and Cruickshank Trust (C.W.), the Scottish Qualifications Authority (E.M.), and the Carnegie Trust for the Universities of Scotland (T.O.).

\end{document}